\documentclass[conference,anonymous]{IEEEtran}

\def\BibTeX{{\rm B\kern-.05em{\sc i\kern-.025em b}\kern-.08em
    T\kern-.1667em\lower.7ex\hbox{E}\kern-.125emX}}

\usepackage{makecell}
\usepackage[utf8]{inputenc}
\usepackage{url}
\usepackage{subfig}
\usepackage{graphicx}
\usepackage{caption}
\usepackage{array}
\usepackage{rotating}
\usepackage{amsmath}
\usepackage{multirow}

\usepackage[official]{eurosym} 

\usepackage{tikz}

\newcommand*\circled[1]{\tikz[baseline=(char.base)]{
            \node[shape=circle,draw,inner sep=0.5pt] (char) {#1};}}

\usepackage{dirtytalk}

\def\CMinit{initial crawl}
\def\CMreap{reappearance crawl}
\def\CMremote{user specific crawl}
\def\CMspoof{spoofing crawl}
\def\CMcont{control crawl}

\def\FeaturesLegend{
CV: Canvas, 
IP: IP address, 
UA: User agent, 
GEO: Geolocation,  
GL: WebGL, 
TZ: Time zone,  
Lang: Accept language, 
DNT: Do Not Track.
}

\def\basiceval#1{\the\numexpr#1\relax}

\def\respwansites{websites including respawning}

\def\ourmethod{cookie respawning with browser fingerprinting}
\def\Ourmethod{Cookie respawning with browser fingerprinting}

\def\websites{$30,000$} 
\def\initcookies{$428,196$} 

\def\reapcookies{$88,470$} 
\def\percreapcookies{$20.66\%$} 
\def\reapwebstes{$18,117$} 
\def\percreapwebstes{$60.39\%$} 

\def\speccookies{$5,144$} 
\def\percspeccookies{$5.81\%$} 
\def\specwebsites{$4,093$} 
\def\percspecwebsites{$22.59\%$} 

\def\excluded{$3,719$} 
\def\respcookies{$1,425$} 
\def\respwebsites{$1,150$}
\def\percrespwebsites{$3.83\%$}

\def\CM13{$CrawlM1_3$}
\def\CM14{$CrawlM1_4$}
\def\CM15{$CrawlM1_5$}

\newif\ifediting

\definecolor{mygray}{gray}{0.2}
\definecolor{mycyan}{rgb}{0.05, 0.5, 0.5}

\ifediting
    \usepackage{comment}
    \usepackage[draft,textsize=footnotesize,textwidth=35mm]{todonotes}
\fi

\def\inlinedremark#1#2#3{        
  \ifediting
        \textcolor{#2}{\textbf{#1: } #3}
    \fi
    }

\def\NBtext#1{\inlinedremark{Nataliia}{orange!100}{#1}}

\def\REMARK#1#2{        
  \ifediting
    \begin{center}
    \noindent\fbox{
        \begin{minipage}[b]{0.4\textwidth}
            \textrm{{[#1]: #2}}
        \end{minipage}}
    \end{center}
    \fi
    }

\def\NB#1{\textbf{\REMARK{Nataliia}{{#1}}}}

\def\IF#1{\textbf{\REMARK{Imane}{{#1}}}}

\def\SHORTEN{\vspace*{-0.5cm}}

\title{Did I delete my cookies?
Cookies respawning with browser fingerprinting. }

\author{
  Imane Fouad\\
  \textit{Inria, France}
    \and
    Cristiana Santos\\
    \textit{Utrecht University}
  \and
  Arnaud Legout\\
  \textit{Inria, France}
    \and
  Nataliia Bielova\\
  \textit{Inria, France}
}

\begin{document}

\maketitle



\begin{abstract}

Stateful and stateless web tracking gathered 
much attention in the last decade, however they were always measured separately.
To the best of our knowledge, 
our study is the first to detect and measure
cookie respawning with browser and machine fingerprinting.
We develop a detection methodology that allows us to detect cookies dependency on browser and machine features. 

Our results  show that {\respwebsites} out of the top {\websites} Alexa websites deploy this tracking mechanism.
We further 
uncover 
how domains collaborate
to respawn cookies through fingerprinting.
We find out that 
this 
technique can be used to track users across websites even when third-party cookies are deprecated. 
Together with a legal scholar, 
we conclude that 
\ourmethod\ 
lacks legal interpretation under
the GDPR and the ePrivacy directive, 
but its use in practice may breach them, 
thus subjecting it to fines up to 20 million \euro{}.

\end{abstract}


\section{Introduction}

In the last decades, the usage of the web  has considerably increased, 
along with the 
 web browsers sophistication.
In parallel, numerous companies built their business models on profiling and tracking web users. 
Therefore, browsers evolution
does not only provide a better user experience, but also allows the emergence 
of new tracking techniques exploited by 
companies to collect users' data. 
There are two main categories of tracking techniques:
stateful and stateless.

\textit{Stateful tracking} is a standard technique that 
 relies on browser storage such as cookies \cite{Roes-etal-12-NSDI,Acar-etal-14-CCS, Ayen-etal-11-Flash,Engl-etal-15-WWW}.
%
Trackers store  a unique identifier in the cookie and later use
it to recognize a user and track her activity across, possibly, different websites.
The simplest way to protect from such tracking is to delete 
the unique identifier by, e.g., cleaning the cookie storage.
However, trackers can recreate deleted cookies using a technique called \emph{cookie respawning} to track users.  
For instance, a tracker can use multiple browser storages 
that store identifiers, in addition to the cookie storage, such as the HTML5 localStorage~\cite{Ayen-etal-11-Flash}.
Consequently, even if the user cleans the cookie storage, the tracker can still recreate cookies using other  storages~\cite{Solt-etal-09-Flash, Ayen-etal-11-Flash,Acar-etal-14-CCS,Roes-etal-12-NSDI}.

\textit{Stateless tracking} allows to track a user without storing identifiers in her browser storage.
Using \emph{browser fingerprinting}~\cite{Niki-etal-13-SP,Cao-etal-17-NDSS,Engl-Nara-16-CCS,Acar-etal-13-CCS,Lape-etal-20-TWEB,Iqba-IEEE-21},
trackers can identify a user through a combination of the user's browser and machine features, such as the user agent or 
IP address.
%
Whereas it is hard to prevent it, browser fingerprinting is not stable over time.
Vastel et al.~\cite{Vast-EUROSP-2018} showed  that fingerprints change frequently: out of 1,905 studied browser instances, 50\%  changed their fingerprints in less than 5 days, and 80\% in less than 10 days.
This instability is caused either by automatic triggers such as software updates or by changes in the user's context such as travelling to a different timezone. 

In summary, 
stateful tracking is a stable way to track a user until she cleans cookies and other browser storages.
Stateless tracking is not stable over time, but does not require any storage and can't be easily stopped by the user.
%
So given that each technique is not perfect, \emph{how can a tracker take advantage of the  best of the two worlds?} 
%
%
The tracker can first use a browser fingerprint to create an identifier and  store it in the browser's cookie.
In this way, even if a user cleans this cookie, the identifier can be recreated with a browser fingerprint. 
Moreover, even if the fingerprint changes over time, the identifier stored in the  cookie can help to match the new fingerprint with the old fingerprint of the same user.
We refer to this tracking technique as \textit{\ourmethod}, that ensures continuous tracking even 
if all cookies are deleted or 
if the fingerprint changes over time. 

Several studies measured the prevalence of 
stateful~\cite{Acar-etal-14-CCS,Roes-etal-12-NSDI,Engl-etal-15-WWW} or stateless~\cite{Niki-etal-13-SP,Cao-etal-17-NDSS,Engl-Nara-16-CCS,Acar-etal-13-CCS} tracking techniques separately. 
However, to the best of our knowledge, \emph{we are the first to study how trackers profit from both stateful and stateless techniques} by combining them. 


The aim of this paper is to  
propose a robust methodology to 
detect and measure the prevalence of {\ourmethod}, followed by a technical and legal analysis of the privacy implications of this tracking technique. 
%
In this paper, we make the following contributions.
\begin{enumerate}
    \item \textbf{We designed a robust method to identify which features are used to respawn a cookie.}
    Our contribution lays in the design of a method to automatically identify the set of fingerprinting features used to generate a cookie, hence, to conclude which user information is collected.
    We additional perform a permutation test (N=10,000, p$<$0.05)) to provide certainty on the dependency between the features and the cookies.
    
    \item \textbf{We make the first study of {\ourmethod}.}
    We show that the stateful and stateless tracking techniques 
    that were studied separately are, in fact, actively used together by trackers. 
%
We found that {\respwebsites} (3.83\%)  of 
the Alexa top {\websites} websites use {\ourmethod}.  

    \item \textbf{We identify who is responsible of {\ourmethod}.}
    We made a detailed study of the responsibility delegation of {\ourmethod}. We show that 
    multiple actors collaborate to 
    access user features, set and own the cookies: 
    we uncovered collaborations between 35 distinct domains that together respawn 67 different cookies.
    \item \textbf{We show that {\ourmethod} is highly deployed in  popular websites.}
     {\Ourmethod} is also happening  on websites from different categories including highly sensitive ones such as adult websites.
     
     \item \textbf{We show that {\ourmethod} lacks legal interpretation and its use, in practice, violates the GDPR and the ePrivacy directive.} 
     We are the first to assess the  legal consequences of this practice together with a legal expert co-author.
     Despite the intrusiveness of this practice, it has been overlooked in the EU Data Protection Law
     and it is not researched in  legal scholarship, nor audited by supervisory authorities.

\end{enumerate}

\section{Background}
 \label{sec:background}
 
 \subsection{Scope of cookies: host and owner}
 \label{sec:back-cookie-host-owner}
In this paper, we make a distinction between the notion of cookie \emph{host} and cookie \emph{owner}.
When a cookie is stored in the browser, it's identified by a tuple ({\em host}, key, value). If the cookie is set via an HTTP(S) response header, then the {\em host} of the cookie represents a domain that set the cookie. 
However, when the cookie is set programmatically via a JavaScript script included in the website, the script gets executed in the context, or ``origin" where it is included. Due to the Same Origin Policy (SOP)~\cite{SameOriginPolicy}, the {\em host} of a cookie set by the script is the origin of the execution context of the script, and not the domain that contains the script.
Given a cookie stored in the browser with its ({\em host}, key, value), when a browser sends a request to a domain, it attaches a cookie to the request if the cookie {\em host} matches the domain or the subdomain of the request~\cite{MDN-HTTP-cookies}.

A cookie {\em owner} is the responsible of setting the cookie. It is either a domain that sets a cookie via HTTP(S) response header (and in this case, matches with the cookie {\em host}), or the domain that hosts a script that sets the cookie programmatically (generally speaking, here the owner is different from host). 
For example, \texttt{site.com} is a  
website that includes a third party script from \texttt{tracker.com}. After loading, the script sets a cookie in the context of the visited website \texttt{site.com}. In this case, the cookie owner is  \texttt{tracker.com}, but the cookie host is \texttt{site.com}.

 \subsection{Web Tracking Technologies}
 \textbf{Cookie-based tracking.}
 Websites are composed of  first party content and numerous third-party content,
such as  advertisements, web analytic scripts, social widgets, or images. 
Following the standard naming~\cite{Lern-etal-16-USENIX}, for a given website we distinguish two kinds of domains: 
 \emph{first-party} domain that is the domain of the website, and 
\emph{third-party} domains that are domains of the \emph{third-party} content served on the website. 

Using HTTP request (or response), any content of the webpage can set (or receive) cookies.
Additionally, cookies can be set programmatically via an included JavaScript library. Every cookie is stored in the browser with an associated domain and path, so that every new HTTP request sent to the same domain and path gets a cookie associated thereto attached to the request. 
First-party cookies set by first-party domains 
are capable to  track users \emph{within the same website}.
Third party cookies set by third-party domains allow third parties to  track users \emph{cross-websites}~\cite{Roes-etal-12-NSDI}.

 \textbf{Browser fingerprinting.}
Browser fingerprinting is a stateless tracking technique that provides the ability to identify and track users without using their browser storage~\cite{Ecke-10-PET,Acar-etal-13-CCS,Cao-etal-17-NDSS,Nas-18-ISC}, unlike cookie-based tracking. 
When a user visits a web page that includes a fingerprinting script, this script will return to a fingerprinter server a list of features composed of user's browser and machine characteristics, such as user agent or time zone.
The trackers use these collected features to build a unique identifier.

\textbf{Cookie respawning.}
Cookie respawning is the process of automatically recreating a cookie deleted by the user (usually by cleaning the cookie storage). 
Several techniques can be used to respawn a cookie. While  related works focused on exploiting another browser storage (e.g., the HTML5 local storage) that duplicates the information contained in the cookie, in this work, we focus on the usage of a browser fingerprint to recreate a cookie.  Section~\ref{sec:methodology} describes how a tracker can exploit a browser fingerprint to respawn a cookie. 

\section{Related work}  
\label{sec:relatedwork}

Cookie based  tracking is a classical tracking technique which has been widely studied in the past decade~\cite{Solt-etal-09-Flash, Roes-etal-12-NSDI,Ayen-etal-11-Flash,Olej-etal-14-NDSS,Engl-Nara-16-CCS,
  Ecke-10-PET,Acar-etal-13-CCS,Niki-etal-13-SP,Lern-etal-16-USENIX,Raza-etal-18-NDSS},   
and is now commonly blocked by modern browsers~\cite{Mozilla,ITP}  and add-ons~\cite{ghostery,disconnect}.
  In this paper, we explore a more sophisticated technique combining cookie based tracking with fingerprinting. 

In 2010, the Panopticlick study showed that fingerprints can be potentially used for web tracking~\cite{Ecke-10-PET}.
Following this study, several fingerprinting tracking techniques were discovered.
Acar et al. studied canvas based fingerprinting~\cite{Acar-etal-13-CCS}.
Englehardt et al. 
presented a new fingerprinting technique based on the AudioContext API.
Cao et al. presented a fingerprinting study mainly based on hardware features including WebGL~\cite{Cao-etal-17-NDSS}.
Al-Fannah et al. studied fingerprinting in Majestic top 10,000 websites~\cite{Nas-18-ISC}.
Solomos et al.~\cite{Solo-NDSS-21} combined browser fingerprinting and favicons caches to identify users.

The term {\em respawning} was first introduced in 2009 by  Soltani et al.~\cite{Solt-etal-09-Flash}.
They showed that trackers are abusing the usage of the Flash cookies in order to respawn or recreate the removed HTTP cookies. This work attracted general audience attention~\cite{media, media2} and triggered lawsuits~\cite{legal1, legal2}. 
Following Soltani et al. work, other studies started analyzing the usage of other storages for respawning such as ETags and localStorage~\cite{Ayen-etal-11-Flash}.
Sorensen studied the usage of  browser cache in cookie respawning~\cite{Sore-IEEE-13}.
Acar et al. automated the detection of cookie respawning and found that IndexedDB can be used  to respawn cookies as well~\cite{Acar-etal-13-CCS}.
Roesner et al.  showed that cookies can be respawned from  local and Flash storages~\cite{Roes-etal-12-NSDI}.

Laperdrix et al.~\cite{Lape-etal-20-TWEB} surveyed recent advancement in measurement and detection of browser fingerprinting. The survey mentions~\cite[\textsection 5.1]{Lape-etal-20-TWEB}  that browser fingerprint together with IP address can be used to regenerate deleted cookies, however no previous work studied this phenomena.
 
{\em Unlike previous works that studied the usage of browser storages to respawn cookies, or measured fingerprinting independently, our study analyzes the usage of fingerprinting to respawn cookies. }
\section{Methodology}
\label{sec:methodology}
\IF{\#134A: Definition of fingerprinting}

When a user  visits a web page with some content located on a tracker's server, the user's browser sends an HTTP(s) request to the server to fetch this content. This request contains several HTTP headers, such as user agent, and an IP address that tracker's server receives \emph{passively}. 
We refer to such information as \emph{passive features}.
To collect additional information, the tracker can include in the visited web page a script that gets executed on the user's browser. The script retrieves multiple browser and machine information, such as the time zone, and sends them to a server of the remote tracker. We refer to such information as \emph{active features}. 
In the following, we define a \textit{browser fingerprint} as the set of active and passive features accessed by the tracker.

We say that a tracker \emph{respawns a cookie} when it recreates the exact same cookie after the user revisit the website in a clean browser.

\subsection{How can trackers benefit from a combination of cookies and browser fingerprint?}

To benefit from both techniques, the tracker can first use a browser fingerprint to create an identifier and  store it in the browser's cookie.
In this way, even if a user cleans this cookie, the identifier can be recreated with a browser fingerprint. 
Moreover, even if the fingerprint changes over time, the identifier stored in the  cookie can help to match the new fingerprint with the old fingerprint of the same user. We explain these scenarios and benefits in details below.

\begin{figure*}[]

\centering

\subfloat[{\Ourmethod} tacking mechanism]{%
\includegraphics[width=0.5\textwidth]{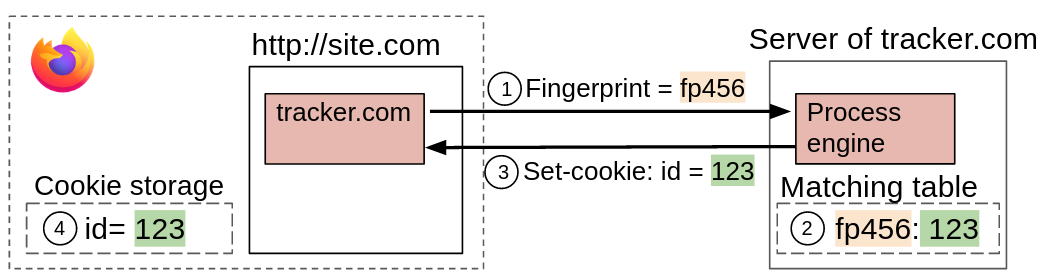}%
}
\subfloat[Recreation of cookies using browser fingerprint]{%
 \includegraphics[width=0.5\textwidth]{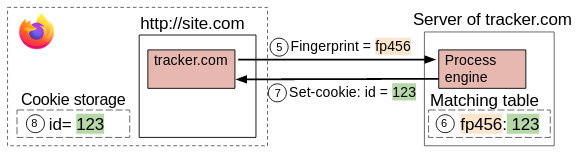}%
}

\subfloat[Usage of cookies to ensure fingerprint stability]{%
 \includegraphics[width=0.5\textwidth]{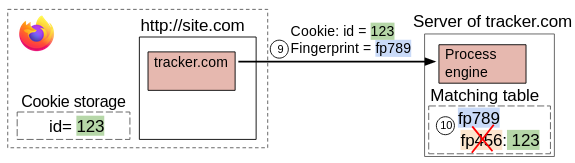}%
}
\caption{\textbf{{\Ourmethod} tracking technique.} (a) { (step 1) The tracker
  receives user's features, 
(step 2) then stores a fingerprint \textit{fp456} associated with the features and generates a corresponding cookie \textit{123}.
(step 3) Next, the tracker sets the cookie in the user's browser.
 (step 4) As a result an identifier is stored in the browser cookie storage.
  (b) When the user cleans her browser and revisit the website, (step 5) the tracker receives the fingerprint \textit{fp456}, (step 6) extracts the corresponding cookie from the matching table, (step 7) and re-sets it in the user's browser. (step 8) As a result, the cookie \textit{123} is recreated in the user's browser. (c) The fingerprint is not stable over time, (step 9) thus the user fingerprint might change. (step 10) The tracker can use the  cookie received with the fingerprint to update the latest on the server side. } }
\label{fig:tracking}
\end{figure*}

Figure~\ref{fig:tracking}(a) shows that the tracker first receives a set of user's active and{\textbackslash}or passive features (step \circled{1}). In step \circled{2}, the tracker generates an identifier from the received features, that it might store on the server's matching table.
The tracker then stores the created identifier in the user's browser cookie, either via the \texttt{Set-cookie} header (step \circled{3}) or programmatically via JavaScript (not shown in Figure~\ref{fig:tracking}(a)). As a result, an identifier is stored in the browser's cookie database (step \circled{4}).

Figures~\ref{fig:tracking}(b) shows what happens when the user does not have a  cookie \textit{123} in her browser, however the fingerprint \textit{fp456} remains the same. In this case, the fingerprint \textit{fp456} is sent to the server of \texttt{tracker.com} (step \circled{5}), and it allows the tracker to match the known fingerprint and the cookie previously set for this user (step \circled{6}). As a result, the tracker is able to set again the same cookie \textit{123}, previously deleted by the user (step \circled{7}). This allows the tracker to respawn deleted user cookies with browser fingerprinting and continue tracking her via such cookies (step \circled{8}).

Figure~\ref{fig:tracking} (c) presents the consequences of \ourmethod. 
When the browser fingerprint of the user is updated from \textit{fp456} to \textit{fp789}, the server of \texttt{tracker.com} receives an old cookie \textit{123} with a new fingerprint \textit{fp789} (step \circled{9}). The cookie \textit{123} helps the server to recognize the user's browser and update the corresponding record in the matching table and substitute a fingerprint \textit{fp456} to \textit{fp789} associated to cookie \textit{123} (step \circled{10}). This allows the tracker to match different fingerprints of the same user, given that fingerprinting is not stable over time.

As a result, \ourmethod\ allows trackers to respawn deleted cookies, and also to link different browser fingerprints of the same user. This makes the tracking robust to either cookie deletion or fingerprint change. Only in case the browser fingerprint changes and the cookie is deleted at the same time, the tracker will not be able to recognize the user and hence to continue tracking this user.

\emph{In this paper, we propose a robust methodology to detect the mechanisms presented in Figures~\ref{fig:tracking} (a) and (b)}.
In this section, 
we first  introduce our methodology to crawl  Alexa top {\websites} websites (Section \ref{sec:data}). 
Next, we present our method to detect {\ourmethod}
(Section~\ref{methodo-detection}). 
Then, we describe the fingerprinting features used in our study and spoofing techniques (Section~\ref{sec:spoofing}). Finally, we list the limitations of our methodology (Section~\ref{sec:limitation}).

\subsection{Measurement setup}
\label{sec:data}

We performed passive web measurement on March 2021 of the Alexa top {\websites} websites extracted on March 2020\footnote{We made this list of websites publicly available~\cite{visitedAlexa}.}. 
All measurements are performed using the OpenWPM platform~\cite{OpenWPM} on the Firefox browser. 
OpenWPM provides browser automation by converting high-level commands into automated browser actions.
We used two machines to perform the crawls in our study.
The versions of OpenWPM and Firefox, the time period of the crawl, and the characteristics of the two machines used in this study are presented in  Table~\ref{tab:characteristics} of the Appendix (Section \ref{sec:appendChara}).

We used different characteristics with  two machines so that they appear as different users, as done by previous works~\cite{fou-IWPE-2020,Acar-etal-14-CCS,Engl-etal-15-WWW,Engl-Nara-16-CCS}. 
Ideally, we would have used two distinct machines with different locations to detect user specific cookies,
however, both machine A and machine B are located in France.
Hence, to change  the Machine B geolocation, we spoofed the
parameters  latitude  and  longitude  by  modifying the  
value of \texttt{geo.wifi.uri} advanced preference in the browser and point it to Alaska.

All our crawls are based on the notion of \textit{stateless crawling instances}. We define a stateless crawling instance of a website \textit{X} as follows:
(1) we visit the home page of the website \textit{X} and keep the page open until all content is loaded to capture all cookies stored (we set the timeout for loading the page to $90$s), 
(2) we  clear the profile by removing the Firefox profile directory that includes all cookies and browser storages. 
The rational behind 
the stateless crawling instance is to 
ensure that we do not keep any state in the browser between two crawling instances. This guarantees that respawned cookies do not get restored from other browser storages.

\begin{table}[]
    \centering
    \begin{tabular}{|p{3 cm}|p{5 cm}|}
    \hline
          \textbf{JavaScript calls} & \textbf{API} \\ \hline
         HTML5 Canvas &  HTMLCanvasElement, CanvasRenderingContext2D  \\ \hline
         HTML5 WebRTC &  RTCPeerConntection \\ \hline
         HTML5 Audio & AudioContext \\ \hline
         Plugin access &  Navigator.plugins \\ \hline
         MIMEType access &   Navigator.mimeTypes \\ \hline

         Navigator properties &  window.navigator \\ \hline
         Window properties & Window.screen, Window.Storage, window.localStorage, window.sessionStorage, and window.name  \\ \hline
    \end{tabular}
    \caption{\textbf{Recorded JavaScript calls.}}
    \label{tab:calls}
            \vspace{-.5 cm}
\end{table}{}

We perform stateless crawling instances of the Alexa top {\websites} websites
and for \textit{each stateless crawling instance}, we extract the following 
from the information automatically collected during the crawls by OpenWPM:  
\begin{enumerate}
    \item For each HTTP request: the requested URL, the HTTP header.
    \item For each HTTP response: the response URL, the HTTP status code, the HTTP header.
    \item All JavaScript method calls described in Table~\ref{tab:calls}.
    \item All cookies set both by JavaScript and via HTTP Responses. On these collected cookies, we perform the following filtering as shown in Figure~\ref{fig:seqe}: first, we select cookies recreated after cleaning the cookies database;
    second, we filter out cookies that are not user-specific;
finally, we filter out cookies that are not respawn with studied features (Section~\ref{methodo-detection}). 
\end{enumerate}


\subsection{Detecting {\ourmethod} with  sequential crawling}
\label{methodo-detection}

\begin{figure*}[!h]
    \centering
    \includegraphics[width=1\textwidth]{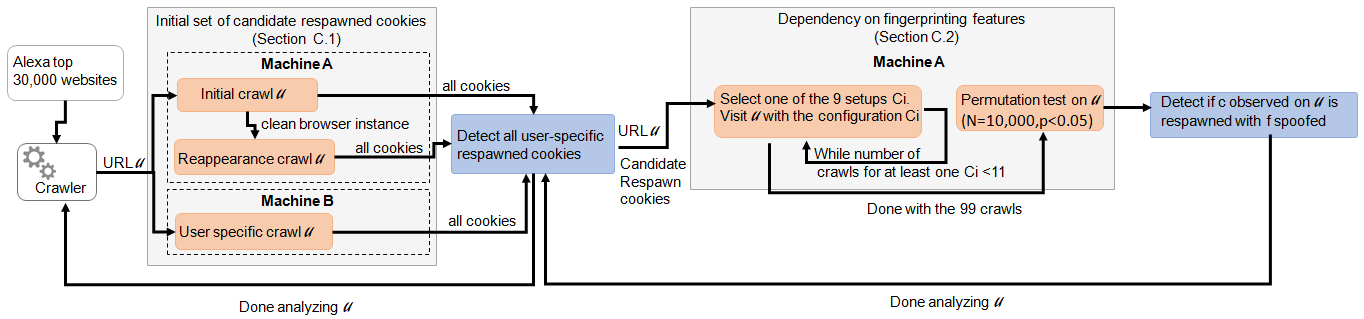}
    \caption{\textbf{Sequential crawling of {\websites} top Alexa websites to identify {\ourmethod}.} \normalfont{ {For each website, 
    we perform an  \textit{{\CMinit}} from machine A and,  a \textit{\CMremote} from machine B to detect machine unrelated cookies.
     After \textit{{\CMinit}} finishes, we start a  \textit{\CMreap} from machine A to detect reappearance of cookies. 
    Using \textit{{\CMinit}},  \textit{\CMremote}, and \textit{\CMreap} we detect \textit{user-specific} cookies that reappear in \textit{\CMreap}, but not in \textit{\CMremote}. 
    For such cookies,
     we randomly chose one configuration $C_i$: either spoof one feature at a time or to set all features to initial value. We perform 99 stateless crawls (11 \textit{\CMspoof}s per feature and 11  \textit{\CMcont}s where the studied features are unspoofed). 
    Finally, we perform a permutation test for each feature (N=10,000),  and we consider that the cookie is \textit{feature dependent} if the resulting p-value $<$ 0.05. All these steps are discussed in  Section~\ref{methodo-detection}.}}}
    \label{fig:seqe}
\end{figure*}

Figure~\ref{fig:seqe} presents our sequential crawling methodology that detects which fingerprinting features are used to respawn cookies. Our method consists of two main steps 
 explained in this section:
\begin{itemize}
    \item {\bf Create the initial set of candidate respawned cookies:} we identify candidate respawned cookies by collecting all cookies that get respawned in a clean browsing instance, and we remove cookies that are not user-specific. 
    \item {\bf Identify dependency of each respawned cookie on each fingerprinting feature:} 
    we spoof each 
    feature independently to detect whether the value of a respawned cookie has changed when the feature is spoofed. 
    We perform a permutation test ($N=10,000$, $p<0.05$) to add statistical evidence on the dependency between a feature and the respawned cookie.
\end{itemize}

\subsubsection{Creation of the initial set of candidate respawned cookies} 
\label{sec:specific}

To build the initial set of candidate respawned cookies, we perform two stateless crawling instances from machine A as described in Figure~\ref{fig:seqe}  (see \textit{{\CMinit}} and \textit{\CMreap}). Via these two crawls, we ensure that all browser storages are cleaned and the only possible way for cookies to be respawn is with browser fingerprinting.

We define a cookie as the tuple \textit{(host, key, value)} where \textit{host} is the domain that can access the cookie.
To create the set of candidate respawned cookies, 
we only collect cookies that 
appear in both the \textit{{\CMinit}} and \textit{\CMreap} when visiting the same website in the two crawls.
Note that due to our sequential crawling (that is, we visit websites in a sequence), we only consider candidate respawned cookies within the same website.

Previous research~\cite{fou-IWPE-2020,Acar-etal-14-CCS,Engl-etal-15-WWW,Engl-Nara-16-CCS}  
considered that cookies are non specific to the users and hence unlikely to be used for tracking  when their values are identical for several users. 
Therefore, using distinct machines to remove non user-specific cookies became a common method in this research area.
We follow this methodology and remove cookies that are not user-specific from our set of candidate respawned cookies.
To do so, we performed an additional 
\textit{\CMremote}\footnote{Practically, we perform the \textit{\CMinit} and \textit{\CMremote}  in parallel, and the \textit{\CMreap} right after the \textit{\CMinit} completes (Figure~\ref{fig:seqe}).} from a different machine B that 
appears to trackers as a different user.  
%
It's important that machines A and B have different fingerprinting features (see Table~\ref{tab:characteristics} of the Appendix) 
to avoid wrong categorization of cookies that depend on these features as non user-specific.

We hence remove the following cookies from the candidate set of respawned cookies and keep only user-specific cookies:
\begin{itemize}
    \item a cookie \textit{(host, key, value)} if it appears on both the \textit{{\CMinit}} on machine A and \textit{\CMremote} on machine B with the same \textit{host}, \textit{key}, and \textit{value}.
    \item a cookie \textit{(host, key, value)}  if a cookie with the same \textit{host} and \textit{key} is not present in a \textit{\CMremote}.  We  adopt a conservative strategy to remove such cookies because we do not have a proof that such cookies are user-specific.
\end{itemize}
Our robust deletion method for cookies that are not user-specific or do not re-appear in a \textit{\CMremote} allows us to ensure that only user-specific cookies are further analysed.

\subsubsection{Identifying dependency of each respawned cookie on each fingerprinting feature.}
\IF{\#134B: statistical test}
\label{sec:feature}
The set of candidate respawned cookies contains cookies that are both user-specific and respawn when crawled a second time after we used a new browser instance with a cleaned browser storage. 
Therefore, cookies in this set are very likely to be  respawned with the use of browser fingerprinting. 
To detect which fingerprinting features are used to respawn the collected cookies, we performed the following steps. We first identified 8 fingerprinting features from previous research (see more details on the choice of features and methods to spoof them in Section~\ref{sec:spoofing}). 
Then,
\def\website{\textit{u}}
for each website \website\ where we have at least one candidate respawned cookie, we perform 99 crawls, 11 spoofing crawls per studied fingerprinting feature \textit{f}, and 11 crawls with all features set to their initial values (as in \textit{\CMinit}) that we refer to as \textit{control crawls}.
In each of the total 88 spoofing crawls, 
we first spoof the feature \textit{f} 
and perform a stateless \textit{\CMspoof} 
of the website \website.
For  each user-specific respawned cookie from the candidate set, we perform the following algorithm.
\begin{itemize}
    \item For each of the 99 crawls, we label the cookie as respawned if the cookie's {\em host} and {\em key} are identical but {\em value} are different from the {\CMinit}. As a result we get 11 observations for each  configuration (either one of the 8 features is spoofed or no feature spoofed.)
    \item For every feature, we perform a permutation test with the 11 observations from the \textit{control crawls}  using  10,000 permutations. The statistical test assess  the difference of the probability to have the cookie respawned between the feature crawls and the control crawls. 
    \item We consider that the cookie is \textit{feature dependent} if the p-value for the test statistic is lower than 0.05.
\end{itemize}

\subsection{Selection of fingerprinting features and spoofing techniques}
\label{sec:spoofing}
%

To achieve a high uniqueness of an identifier built from a browser fingerprint, trackers use a combination of both passive and active browser and machine features. 
Though browser features are useful for fingerprinting, using them alone might be problematic for trackers because the usage of multiple browsers is recommended and common among users~\cite{US-CERT,Berger, Cao-etal-17-NDSS}.
To improve the accuracy of the fingerprint, trackers also use machine related features such as the IP address, or the OS version~\cite{Boda-etal-11-Nordec, Nas-18-ISC}.


Table \ref{tab:sfeatures} presents a full list of studied browser and machine features that we selected based on the most common features in prior works for browser fingerprinting~\cite{Lape-etal-16-SP, Boda-etal-11-Nordec,gomezboix-www-2018, Nas-18-ISC,Cao-etal-17-NDSS, Acar-etal-14-CCS,Das-CCS-2018, Engl-Nara-16-CCS,Mowe-W2SP-12}. 

\begin{table}[]
    \centering
    \begin{tabular}{|p{1.3 cm}|p{4 cm}p{2 cm}|}
\hline
            \multirow{ 4}{*}{\makecell{\textbf{Browser} \\ \textbf{features}}} 
            & Accept language~\cite{Lape-etal-16-SP, Nas-18-ISC} & Active/Passive   \\ 
        &  Geolocation~\cite{Nas-18-ISC}  & Active \\
        & User agent~\cite{Lape-etal-16-SP, Boda-etal-11-Nordec,gomezboix-www-2018, Nas-18-ISC} & Active/Passive  \\ 
        & Do not track~\cite{Lape-etal-16-SP, gomezboix-www-2018}   & Active/Passive \\ 
           \hline
           \hline
 \multirow{ 4}{*}{\makecell{\textbf{Machine} \\ \textbf{features}}} & WebGL~\cite{Lape-etal-16-SP, Cao-etal-17-NDSS,gomezboix-www-2018,Nas-18-ISC,Mowe-W2SP-12} & Active \\ 
 & Canvas~\cite{Lape-etal-16-SP, Cao-etal-17-NDSS, gomezboix-www-2018, Acar-etal-14-CCS,Das-CCS-2018, Engl-Nara-16-CCS,Mowe-W2SP-12} & Active \\
 & IP address~\cite{Boda-etal-11-Nordec, Nas-18-ISC,Das-CCS-2018,Engl-Nara-16-CCS}&   Passive\\ 
 & Time zone~\cite{Lape-etal-16-SP,Boda-etal-11-Nordec, gomezboix-www-2018} & Active \\ \hline
    \end{tabular}
    \caption{\textbf{Studied fingerprinting features.}}
    \label{tab:sfeatures}
            \vspace{-.4 cm}
\end{table}

We have used two methods to spoof fingerprinting features: 1) via Firefox preferences and 2) add-ons. We have validated that each feature has been properly spoofed on our own testing website with a fingerprinting script  and also by using \textit{whoer} website~\cite{whoer} that verifies the information sent by the user's browser and machine to the web.

\subsubsection{Spoofing using Firefox preferences}
Firefox allows to change its settings in the browser preferences of \textit{{about:config}} page.
With this method,
we spoofed the following features.

 \textbf{User agent.}
The \textit{User-Agent} HTTP header allows the servers to identify the operating system and the browser used by the client.
The  \textit{{\CMinit}} run in  Firefox under Linux (see Table \ref{tab:characteristics} in the Appendix for details).
To spoof the user agent, we changed the \texttt{general.useragent.override} preference in the browser to Internet Explorer under Windows:
(\textit{Mozilla/5.0 (Windows NT 6.1; WOW64; Trident/7.0; AS; rv:11.0) like Gecko}).
We checked the spoofing efficiency on our testing website with an injected script.
The script returns the user agent value using the \texttt{navigator.userAgent} API.
We tested the user agent value returned with the HTTP header  using the \textit{whoer} website. 
We found that the user agent value was spoofed both in JavaScript calls and HTTP headers.

\textbf{Geolocation.} 
The geolocation is used to identify the user's physical location.
The \textit{{\CMinit}} has as location  \textit{Cote d'Azur, France}.
We spoofed  the geolocation parameters latitude and longitude by modifying 
the value of \texttt{geo.wifi.uri} preference in the browser  and point it to the \textit{Time Square, US} ("lat": 40.7590, "lng": -73.9845).
We validated the spoofing efficiency using a script call to \texttt{navigator.geolocation} API.

\textbf{WebGL.} 
The WebGL API is used to give information on the device GPU. 
In our study, we focus on the WebGL renderer attribute that precises the name of the model of the GPU.
We spoofed the WebGL renderer using the \texttt{webgl.render-string-override} preference in the browser.
We changed the value of WebGL renderer to \textit{GeForce GTX 650 Ti/PCIe/SSE2}.
To retrieve information about the graphic driver and read the WebGL renderer value,
we used the \texttt{WEBGL\_debug\_renderer\_info} add-on. We validated the WebGL spoofing efficiency by using  the add-on in our customized website.

 \textbf{Do Not Track.} 
The Do Not Track (DNT) header indicates user's tracking preference.
The user can express that she doesn't want to get tracked by setting the DNT to True.
In the  \textit{{\CMinit}}, the DNT was set to \textit{null}.
We enabled the Do Not Track header, and we set it to True using the  \textit{privacy.donottrackheader.enabled} preference.
We validated that the DNT  returned value in the HTTP header is set to True using the \textit{whoer} website.

\subsubsection{Spoofing using browser add-ons}
The browser preferences do not provide a spoofing mechanism for all fingerprinting features.
We used browser add-ons to spoof such features.

{\textbf{Canvas. }}
The  HTML canvas element is used to draw graphics on a web page.
The difference in  font rendering, smoothing, as
well as other  features cause devices to draw images and texts differently.
A fingerprint can exploit this difference  to distinguish users.
We spoofed the canvas by adding a noise that hides the real canvas fingerprint.
To do so, we used the Firefox add-on \textit{Canvas Defender}~\cite{Canvas}.
To test the add-on efficiency, 
we built a customized website where we inject a canvas fingerprinting script.
The script first draws on the user's browser. Next, the script calls the Canvas API
\textit{ToDataURL} method to get the canvas in dataURL format and returns its hashed value.
This hashed value can then be  used as a fingerprint.
To evaluate  the add-on efficiency against the canvas fingerprinting,
we revisited the customized website  and compared the rendered canvas fingerprint.
We found that the returned canvas hashed values were different upon every visit.

{\textbf{IP address.} } 
We run the  \textit{{\CMinit}} with an IP address pointing to France.
We spoofed the IP address using a VPN add-on called \textit{Browsec VPN}~\cite{VPN}. 
We used a static IP address pointing to the Netherlands.
Consequently, the spoofed IP address remain constant during the runs of spoofed crawls.
We checked that the IP address changed using the \textit{whoer} website.

{\textbf{Time zone.}}
We launched the  \textit{{\CMinit}} with \textit{Paris UTC/GMT +1} timezone.
We spoofed the  timezone to 
\textit{America/Adak (UTC-10)} using the Chameleon add-on~\cite{cham}.

\textbf{Accept-language. }
The \emph{Accept-language} header precises which languages the user prefer.
We  used English as \emph{Accept-language} in \textit{{\CMinit}}.
We spoofed the accept-language header using the Chameleon add-on~\cite{cham} to Arabic.
We checked that it was properly spoofed using the \textit{whoer} website.

\IF{\#134A: Selection of fingerprinting attributes}
\subsection{Limitations}
\label{sec:limitation}

Spoofing features and  implementing the spoofing solution with the OpenWPM crawler requires substantial engineering effort.
Therefore, we limit our study to 8 browser features that are commonly used by previous works and that can be spoofed either directly using browser settings, or using the add-on (Canvas Defender, Browsec VPN, and Chameleon) that we successfully run with OpenWPM.
Consequently, cookies respawned using other features are excluded from this study.
The number of excluded cookies is  2,976 (see Section~\ref{sec:general}.
This is a limitation that does not impact the main goal of our study, as we do not intend to be exhaustive in the identification of respawned cookies, 
but we aim to understand and describe the mechanisms behind respawning, and propose a robust methodology to detect features that are used by trackers to respawn cookies.

Given that we spoof one feature at a time, we may introduce inconsistency between different  features. For example, when we spoof the geolocation API, we don't modify the time zone or the IP address. This 
method doesn't invalidate our results because we detect dependency 
on each feature separately. Nevertheless, we may miss trackers that modify their behaviour 
when some features 
are spoofed.

Non user-specific cookies are not intrusive for the user's privacy because they are identical among different users. We are aware that the cookies we classify as 
non user-specific
might have been respawned due to features we do not consider.

\section{Results}
\label{sec:overview}
In this section, we present
findings on prevalence of {\ourmethod}, identify responsible parties, and analyze on which type of websites respawning occurs more often. 
Our results are based on Alexa top {\websites} websites where we extracted a total of {\initcookies} cookies. We study the respawning of both first and third party cookies.

\subsection{How common is {\ourmethod}?}
\label{sec:general}

Table~\ref{tab:overview}  presents an overview of the prevalence of {\ourmethod}.
We extracted {\initcookies} cookies from the visited {\websites} websites.
Using the \textit{\CMreap}, 
we extracted a set of cookies 
that did  reappear in the crawl.
As a result, 
we 
obtained a set of {\reapcookies} ({\percreapcookies})  reappearing cookies 
that appear on {\reapwebstes} ({\percreapwebstes}) websites.

\begin{table}[]
    \centering
    \begin{tabular}{|p{2.2 cm}|p {.9 cm}  p {1.4 cm} p {1.3 cm} p {1.2 cm}|}
    \hline
     \textbf{Crawls} & \textit{Initial} & \textit{Reappearance} & \textit{User specific} & \textit{Feature dependent} \\
     \hline
     \hline
         Collected cookies  & 428,196  & 88,470 & 5,144 & 1,425 \\
         Occurrence on websites &  30,000 & 18,117  & 4,093 & 1,150 \\ 
         \hline
    \end{tabular}
    \caption{ \textbf{{\Ourmethod} is common on the web.} We detected  1,425 respawned cookies that appear on 1,150 websites. 
    \normalfont{We define the Initial, Reappearance, User specific crawls and Feature dependent cookies in Section~\ref{methodo-detection}. }}
    \label{tab:overview}
\end{table}

Next, we filtered out 
cookies that are not user-specific -- they appear with the same (host, key, value) on \textit{\CMinit} and \textit{\CMremote} -- and 
cookies that only appear on \textit{\CMinit} 
but not in \textit{\CMremote} (Section~\ref{sec:specific}).
We found that out of {\reapcookies} reappearing cookies, {\speccookies} ({\percspeccookies}) are user specific.
The set of user specific cookies 
is observed on {\specwebsites} ({\percspecwebsites}) websites. 

After filtering out non reappearing cookies and keeping only user specific cookies,
we identified cookies whose value depend on at least one of the studied features following our methodology detailed in Sections~\ref{sec:feature}.
As a result, we extracted {\respcookies} respawned cookies that appear on {\respwebsites} ({\percrespwebsites}) websites.
Out of 
the remaining  {\excluded} set of cookies, $743$ were excluded from the statistical test because they did not appear on the 99 spoofing and control crawls.
The remaining $2,976$ cookies that are user specific and not detected as feature dependent can be respawned via other features that are out of scope of our study.


\noindent
{{\bf Summary. } We found {\respcookies}  cookies respawned using at least one of the studied features. These cookies were respawned in  {\respwebsites} websites that represent {\percrespwebsites} of the visited websites.}

\subsection{Which features are used to respawn cookies?}
\label{sec:resultsFeatures}
In this section, we present the results we obtained from the sequential crawling methodology (Section~\ref{methodo-detection}).
For each of the {\respcookies} respawned cookies, 
we detected features on which the cookie value depends
(see all studied fingerprinting features in  Table~\ref{tab:sfeatures}). 

Given that a cookie can be respawned with several features, 
we consider that a cookie \textit{C}  {\em is respawned with a set of features \textit{F}} if the value of \textit{C} depends on every feature in \textit{F} 
(such detection was done independently for each feature as described in 
 Section~\ref{sec:feature}).



\begin{table}[]
    \centering
    \begin{tabular}{|p{1.3cm}|p{0.4 cm}|p{.4 cm}p{.4 cm}p{.5 cm}|p{.4 cm}p{.4 cm}p{.4 cm}p{.4 cm}|}
    \hline
     & \multicolumn{1}{c|}{\textbf{Passive}} & \multicolumn{3}{c|}{\textbf{Active/Passive}} & \multicolumn{4}{c|}{\textbf{Active}} \\ 
     \hline
     \hline
        {Features}  &  \textbf{IP} & \textbf{UA} & \textbf{Lang} & \textbf{DNT} & \textbf{CV} & \textbf{GEO} & \textbf{GL} & \textbf{TZ} \\
        \hline
           {Occurrence} &  672 & 486 & 278 & 277 & 231 & 249 &292 &  310 \\ 

         \hline
    \end{tabular}
    \caption{\textbf{IP address is the most commonly used feature to respawn cookies. }\normalfont{Occurrence: number of times a feature has been used  to respawn a cookie (either independently or in combination with other features). 
    } {\FeaturesLegend}}
    \label{tab:top}
\end{table}

Table~\ref{tab:top} presents the number of times each feature is used to respawn a cookie.
IP address is the most commonly used feature to respawn cookies and is used in respawning of 672 (47.15\%) cookies.
The second most popular feature to respawn cookies is {User-Agent} (UA) -- it is observed with 486 (34.10\%) cookies.
Note that 
features that can be easily collected passively, like IP address and UA, are more frequently used than features that can only be accessed actively, such as Canvas or Geolocation.


\begin{figure}[]
\centering
{\includegraphics[width=0.38\textwidth]{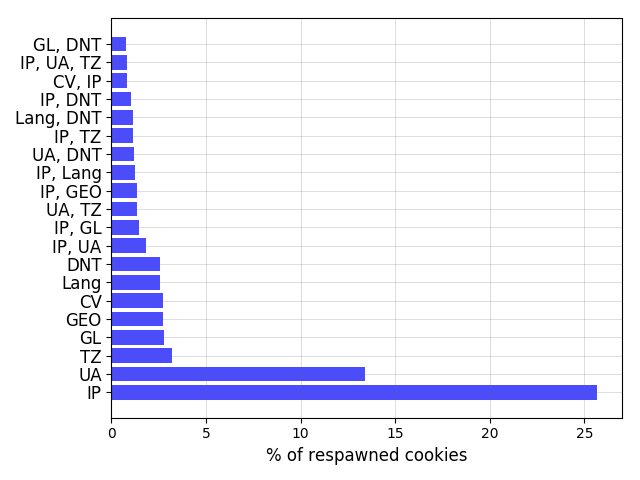}}
\caption{\textbf{Top 20 set of features used to respawn cookies.} \normalfont{IP addresses alone are used to respawn over 25\% of the cookies. \FeaturesLegend}} 
\label{fig:feature} 
\end{figure} 

We found that cookies are usually respawned with a set of different fingerprinting features. In our dataset, cookies are respawned with 184 distinct sets of features.
Figure~\ref{fig:feature} shows 
the sets of features most often used for cookie respawning.
We see that the IP address alone is the most commonly used feature to respawn cookies, and moreover no other set of features is more popular than the IP address alone.

The IP address is used alone to respawn $366$ (25.68\%) cookies.
Mishra et al.~\cite{vika-www-20} studied the stability and uniqueness of the IP address over a duration of 111 days on a dataset of 2,230 users.
They showed that 87\% of participants retain at least one IP address for more than a month. 
Hence, IP addresses are both stable and unique, therefore, they can be used to uniquely identify and track user's activity.
\IF{\#134A: Correlating fingerprinting attributes}
Interestingly, the top-2 sets of features, \{IP\}, and \{UA\}, contain only passive features that are easier to collect. 
Active features are rarely use, 
timezone, the most popular active feature 
for respawning, is 
used alone for  46 (3.23\%) cookies.

\noindent
{{\bf Summary. } We show that trackers use multiple combinations of features to respawn cookies and that the IP address, which is overlooked in a number of fingerprinting studies~\cite{Lape-etal-16-SP, Cao-etal-17-NDSS, gomezboix-www-2018, Acar-etal-14-CCS, Mowe-W2SP-12},  is the most used feature to respawn cookies.} 


\subsection{Discovering owners of respawned cookies}
\label{sec:resp}

 
Cookie respawning opens new opportunities for different companies to collaborate together to track users. 
Usually, the {\em host} of a cookie defines the domain that can access the cookie. We introduce in this paper a notion of cookie {\em owner} that has set the cookie via an HTTP header or programmatically via JavaScript (see Section~\ref{sec:back-cookie-host-owner}). However, additional stakeholders can help to respawn a cookie by serving a fingerprinting script. 
We explore each of these new potential stakeholders in the rest of this section. 




\subsubsection{{Identifying cookie owners}}
\label{sec:owners}

Due to the the Same Origin Policy (SOP)~\cite{SameOriginPolicy}, the domain that is responsible for setting a cookie can be different from the domain that receives it (see Section~\ref{sec:back-cookie-host-owner}).
Therefore, we differentiate two stakeholders: \textit{Owner} -- the domain that is responsible for setting the cookie, 
and \textit{Host} -- the domain that has access to the cookie and to whom the cookie is sent by the browser.
In the following, we define both owner and host as  $2^{nd}$-level TLD domains (such as \texttt{tracker.com}).

It's important to detect the cookie owner -- for instance, in order to block its domain 
via filter lists~\cite{easylist,easyprivacy,disconnect} and prevent cookie-based tracking. Indeed, the notion of cookie owner is often overlooked when the reasoning is only based on the cookie host~\cite{Cahn-WWW-2016}. 
When one cookie owner sets a cookie in the context of several websites (the owner's script can be embedded directly on a visited website or in a third-party \textit{iframe}), the host of this owner's cookie is the context where the cookie is set because of the SOP~\cite{SameOriginPolicy}.
To identify  cookie owners in the context of our paper, we distinguish two cases, as described below.

\IF{\#134A: Identification of JavaScript cookie owner}
{\bf Cookie set by a script.}
\texttt{Document.cookie} property is the standard way for a JavaScript script to set a cookie~\cite{JScookies} programmatically. 
To check whether a cookie is set via JavaScript and to extract its {\em owner} (the domain who serves the script) when crawling a website, 
we
(1)  extract the set of scripts \emph{S} that set a cookie on the website 
using \texttt{document.cookie}, 
(2) for every script in \emph{S}, we extract the set of cookies \texttt{$C$} set by this script, and 
(3) check whether there is an overlap between the set of respawned cookies identified in Section~\ref{sec:general} and in the set \texttt{C}. If it is the case, we conclude that the cookies in the overlap are set via JavaScript, and their {\em owner} is the $2^{nd}$-level TLD domain that served the script.

{\bf Cookie set by HTTP(S) header.}
If the cookie is set by the HTTP(S) \texttt{Set-Cookie} response header, then the {\em owner} of the cookie is the same as its host 
because it  corresponds to the $2^{nd}$-level TLD of the server that set the cookie.

\begin{figure}[]
    \centering
    \includegraphics[width=0.25\textwidth]{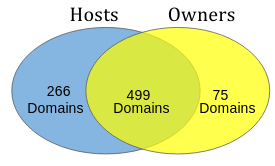}
    \caption{\textbf{Emergence of new domains when considering cookie owners.} \normalfont{The {\respcookies} respawned cookies have 765 distinct hosts and 574 distinct owners.} {The notion of cookie owner allows to identify 75 cookie owner domains that never appear as a cookie host. We also found 266 cookie host domains that are never used to set the cookie.}}
    \label{fig:hoc}
\end{figure}

For each of the {\respcookies} respawned cookies, we identified its {\em owner} depending on how the cookie was set.
Figure~\ref{fig:hoc} shows  domains appearing as host only (left blue part), as owner only (yellow part), or both (middle overlap).
In total, \respcookies\ respawned cookies 
are labeled with 765 distinct hosts, however they were set by 574 distinct owners. 
\IF{\#134A: Effeciency of EasyList/EasyPrivacy or Disconnect to block owners}
Figure~\ref{fig:hoc} also depicts that 
75 domains appear as owners and never as cookies hosts. 
These domains serve JavaScript scripts that set cookies, but  never serve cookies directly via an HTTP(S) response header.
Hence, when only considering cookies hosts, these domains are not detected.
We evaluated the efficiency of disconnect~\cite{disconnect} filter list in detecting these 75  domains. We found that disconnect miss 53 (70.66\%) owners domains.
 We also found that 266 domains that appear as cookie hosts are never identified as cookie owners. 
Cookies associated with these domains were set in the context of the hosts domain because of the SOP, but these domains were never actually responsible of setting these cookies.

\begin{figure}[]
    \centering
    \includegraphics[width=0.38\textwidth]{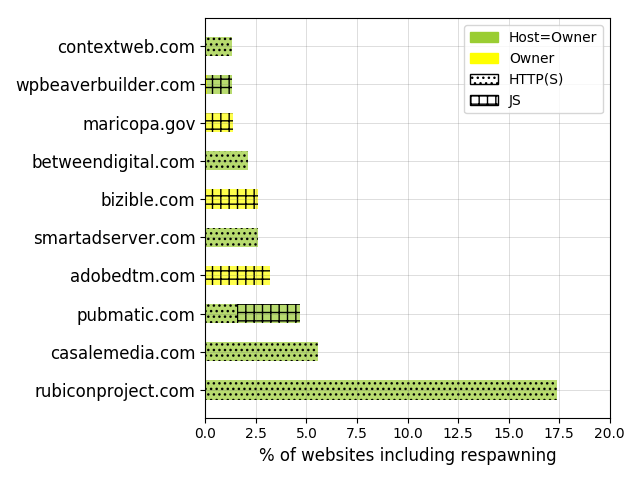}
\caption{\textbf{The top 10 respawned cookies owners.  }
\normalfont{ The bar is green when the domain is both host and owner, and yellow when the domain only appears as owner. 
For each domain, we show when cookies are set via an HTTP(S) header and when they are set via JavaScript.
 When considering cookie owners, new domains are identified such as \texttt{adobedtm.com}.
} }
\label{fig:owners} 
\end{figure}

Figure~\ref{fig:owners} presents the 
top 10 domains responsible for cookie respawning 
that are either cookie hosts, cookie owners, or both. Two domains, \texttt{rubiconproject.com} and \texttt{casalemedia.com}, represent the largest fraction of websites. All cookies served by these two domains are served via HTTP(s). Three out of the top 10 domains are exclusively cookie owners:  \texttt{adobetm.com},  \texttt{bizible.com}, and \texttt{maricopa.gov}. These domains are only setting respawned cookies via JavaScript and never directly through HTTP(S). 
Out of the {\respcookies} respawned cookies, 514 (36.07\%) are set via JavaScript.

\noindent
{{\bf Summary. } 
Previous studies that only looked at the cookie host can miss the trackers responsible for setting the cookies.
In our study, 75 domains could be missed if we only considered cookie hosts. We found that disconnect miss 70.66\% of these domains. Considering cookie owners improves the understanding of the tracking ecosystem.}

\subsubsection{{Identifying scripts used for respawning}}
\label{sec:respawners}
A cookie can be respawned using a set of different features.
These features can be all accessed by a single script or by multiple collaborating scripts as we describe in this section.
To identify the scripts that are responsible of accessing browser or machine features used for respawning a cookie, 
we use the recorded JavaScript  calls described in Table~\ref{tab:calls}.

Every feature can be accessed only actively, only passively or actively and passively (see Table~\ref{tab:sfeatures}).
In this section, we focus {\em only on the active features} collected using the following JavaScript calls: \texttt{window.navigator.geolocation} (to access the Geolocation) and  \texttt{HTMLCanvasElement} (to access the Canvas).
As OpenWPM does not log calls to Time zone and WebGL, we do not consider these active features in this section.
For every respawned cookie $\texttt{C}$,
we identified the set of features $\textit{F}$ used for respawning \texttt{C} as described in Section~\ref{sec:feature}.
To extract the scripts that are responsible of respawning  $\texttt{C}$ via the  set of features $\textit{F}$,
we analyze the features used to respawn each cookie.
If the cookie is respawned with only passive or active/passive features, then no conclusions can be made for both HTTP(S) and JavaScript cookies.
In fact, these features are sent passively, therefore no conclusion can be made on which scripts used them in respawning.

In total, we found that 931 (65.33\%)  
cookies are respawned with only passive or mixed features.
For the remaining 494 cookies depending on active features, 95 (19.23\%)
are only using WebGL or Time Zone that are out of the scope of this study.
In the rest of this section, we consider the 399 respawned cookies that are respawned with only active features for which we can access method calls.

We refer to the set of active features used to respawn a cookie as ${f_a}$. We extract the set of features accessed by every script on the website where the cookie is respawned, and
 distinguish  three cases.

\textbf{1 - The owner of the cookie is suspect to be the responsible of respawning. }
 We identify such cases when (1)  the cookie is set via JavaScript, and all active features ${f_a}$  used to respawn it are  accessed by the owner of the cookie, or (2) the cookie is set via HTTP(S) and a script hosted by the same $2^{nd}$-level TLD  accesses all active features.
    If one of the two cases is validated,  then we suspect that the owner of the cookie is the responsible of respawning it.
    In total, we found 37 (9.27\%) cookies that are respawned by their owners.  Out of these 37 cookies, 17 are set via HTTP(S) and respawned by scripts that belong to the same domain.
    These 37 respawned cookies are owned by 23 distinct owners. Table~\ref{tab:ownerresp} presents the top 4 owners that are suspect to respawn the cookies as well.
    
    \begin{table}[]
    \centering
    \begin{tabular}{|p{3cm} p{2 cm}|}
    \hline
    \textbf{Owner} & \textbf{\# of cookies} \\ \hline
    \hline
    adobedtm.com & 10 \\
    ssl-images-amazon.com & 3 \\
    hdslb.com & 2 \\
    bitmedia.io &  2 \\
    \textit{19 Others} & 20  \\ 
    \hline
    {Total} & 37 \\ \hline 
    \end{tabular}
    \caption{\textbf{Top domains suspect to set and respawn the cookies.}}
    \label{tab:ownerresp}
    \end{table}

We found that \texttt{adobedtm.com}~\cite{adobedtm} (the tag manager owned by adobe) is the top domain that both owns and respawns cookies. Though respawning is not explicitly indicated in their policy, the policy states that they collect browser and machine features.
We didn't find any information regarding  cookies respawned either by \texttt{ssl-images-amazon.com}, \texttt{hdslb.com} or \texttt{bitmedia.io}~\cite{bitmedia}.

 \textbf{2 - The respawning is a result of a potential collaboration between the cookie owner and other scripts. } If the set of active  features used to respawn the cookie are not accessed by the owner, but are accessed by other scripts on the same website, then we suspect that the cookie is potentially a result of collaboration between the owner of the cookie and other scripts on the same website. In this study, we don't assess whether the domains are actively collaborating, or if one domain is leveraging scripts from other domains to glean fingerprint
  information.
In total, we found that 67 (16.79\%) cookies  are suspect to be a result of collaboration between  multiple domains.
The 67 cookies are a result of collaboration of 35 distinct domains.

\begin{table}[]
\centering
\begin{tabular}{|p{2.8cm} p{3.18cm}  p{1.5cm}|}
\hline
\textbf{Owner} & \textbf{Collaborator} & \textbf{\# of cookies} \\ \hline
\hline
\texttt{rubiconproject.com} & \texttt{googlesyndication.com} &  8 \\ 

\texttt{rubiconproject.com} & \texttt{pushpushgo.com} &  3 \\
\texttt{adobedtm.com}& \texttt{morganstanley.com} & 2 \\ \texttt{adobedtm.com}& \texttt{provincial.com} & 2 \\ 
    \hline
    \end{tabular}
    \caption{\textbf{Top domains suspect to collaborate to respawn cookies.} The reported domains are both first- and third- party.}
    \label{tab:collab}
    \end{table}
    
Table~\ref{tab:collab} presents the top domains that are suspect to collaborate in order to respawn cookies.
We define the collaborator as the only domain accessing the features used for respawning the cookie and not accessed by the owner of the cookie.
The top  collaboration involves \texttt{googlesyndication.com} owned by Alphabet (parent company of Google).
\texttt{Googlesyndication.com} is accessing and potentially sharing user's Canvas information\footnote{We will make the data available upon paper acceptance.}.

{ \textbf{3 - The responsible of respawning the cookie are not all known.}}
If not all the active  features used to respawn the cookie are accessed on the website where the cookie is respawned via JavaScript calls, then
we assume that the owner is accessing the features via other means. 
This happens with 295 (73.93\%) cookies. 
In 186  (63.05\%) cookies out of the 295, the owner is not accessing the geolocation API and do access other active features it used for respawning the cookie.
This can potentially be a result of the dependency between geolocation and IP addresses. When we spoof the geolocation to Time Square in the US, we keep an IP address that points to France because we only spoof one feature at time.
Hence, companies may detect this incoherence, and not use the IP address to respawn the cookie, which, in our experiment will be identified as dependency on the geolocation feature.

\noindent
{{\bf Summary. } Identifying the responsible of respawning can prove to be a complex task. While 23 owners respawn cookies themselves, 35 domains collaborate to respawn cookies.}

\subsection{Where does respawning occur?}
\label{sec:where}
In section~\ref{sec:resp}, we studied the domains that are responsible of setting and respawning cookies. 
In this section, 
we analyse on which types of websites respawning occurs.
In the following, we refer to these websites as \textit{{\respwansites}}.
We analyse Alexa ranking distribution and impact of websites category on the usage of cookie respawning, present {\respwansites} that process special categories of data, and  present the geolocation of owners of respawned cookies and websites.

\textbf{Popularity of  {\respwansites}.}
We detected {\respwebsites} websites 
where at least one cookie is respawned.
Table~\ref{tab:ranking} presents the number of {\respwansites} for each Alexa rank interval.
We observe that {\ourmethod} is heavily used on popular websites: out of the top 1k visited websites, 4.9\% are {\respwansites}. This percentage decrease to  3.70\% in less popular websites.

\begin{table}[]
    \centering
    \begin{tabular}{|p{2 cm}  p {2.5 cm } p{2 cm}|}
    \hline
  \textbf{Alexa rank interval} 
  &\textbf{Websites  including respawning} 
  & \textbf{\# of owners} \\ \hline
   \hline
        0 — 1k  &  49 (4.9\%) & 49 \\ 
        1k — 10k & 360 (4\%) &  213\\ 
         10k+ & 741 (3.70\%)  &  382 \\ 
         \hline
     \end{tabular}
    \caption{{\textbf{Popular websites are more likely to include cookie respawning.} \normalfont{Number of owners: presents the total number of distinct respawned cookies owners in the Alexa ranking interval. }}}
    \label{tab:ranking}

\end{table}

\textbf{Categorization of {\respwansites}.} We used the \textit{McAfee service}~\cite{mcafee} to 
categorize the visited websites.
The McAfee uses various technologies and artificial intelligence techniques, such as link crawlers,  and customer logs to categorise websites.
It is used by related works~\cite{Urb-www-20}.
A description of the reported McAfee categories can be found in the McAfee reference guide~\cite{mcafeeCateg}.

We successfully categorized Alexa
29,900 visited websites.
For every category, we present the percent of respawn websites.
We found that the  visited websites belong to 669 categories and the {\respwebsites} {\respwansites} belong to 143 different categories. 

\begin{figure}[]
    \centering
    \includegraphics[width=0.37\textwidth]{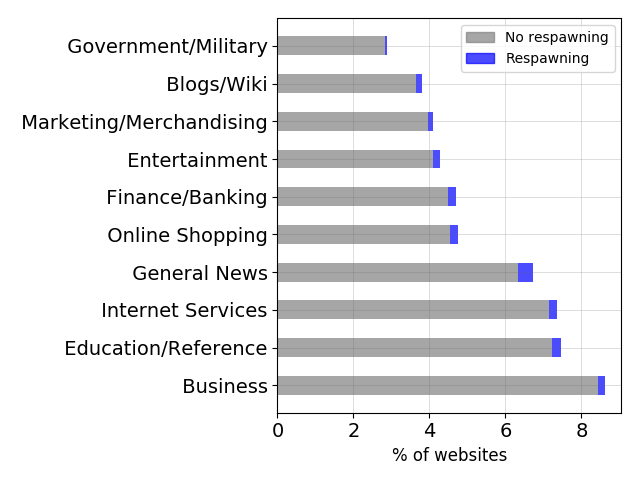}
    \caption{\textbf{General news is the top category including {\ourmethod}.} \normalfont{We consider that a website \textit{U} is including respawning if it contains at least one respawned cookie. The bar is gray when we don't detect respawning in the website, and is blue when we do.} }
    \label{fig:topsites}
    \vspace{-.25 cm}
\end{figure}

Figure~\ref{fig:topsites} gives an overview of the 10 most prominent categories within the Alexa visited websites.
We found that all top 10 categories 
contain  websites that include respawning.
Business is the top websites category, 8.62\% of the visited websites are categorized as business.

Most of {\respwansites} are categorized as \textit{General News}.
Out of the $29,900$ visited websites,  
6.73\%  are categorized as \textit{General News},
and 5.95\% of these \textit{General News} websites contain at least one respawned cookie. 
\textit{General News} is known for using more third parties than other categories~\cite{kir-www}, which can 
be the reason behind the high deployment of respawning in this category of websites.

\textbf{Websites processing special categories of data.}
The GDPR~\cite[Recital 51]{gdpr} stipulates 
that personal data which are particularly sensitive by their nature, merit specific protection, as  their processing could create significant risks to the fundamental rights of users.
Such data include personal data revealing sensitive information such as 
data concerning a natural person’s sex life or sexual orientation~\cite[Article 9]{gdpr}.
Processing such categories of data is \emph{forbidden}, unless allowed by the user's explicit consent~\cite[Article 9(2)]{gdpr}.

We studied tracking via the third-party respawned cookies on websites processing sensitive data.
As a result, we detected $21$ cookies respawned in \textit{Adult} websites  
that are set by $19$ different owners.
The top domain respawning cookies on sensitive websites is \texttt{adtng.com} (no corresponding official website was found for \texttt{adtng.com}). It respawned cookies on 3 different adult websites,
and therefore, 
 can track and link user's activity within adult websites in a persistent way, {\em without explicit consent to legitimize such operation}, rendering such respawning practise unlawful.

\textbf{Geolocation of {\respwansites} and respawned cookies owners.}
Independently of the country of registration of a website, if a website {\em monitors} the behavior of users while they are in the EU, the GDPR applies to such monitoring~\cite[Article 3(2)(b)]{gdpr}.
Notice that any form of web tracking will be deemed as "monitoring", including {\ourmethod}.
Since our experiments simulate users located in France (EU), both EU and non-EU organizations must comply with the GDPR.

We extracted the country  of registration of the  owners of respawned cookies and the websites including them using the \emph{whois library}~\cite{whois}.
We successfully identified the country of registration of 362 (63.07\%) out of 574 total distinct owners, and 670 (58.26\%) out of 1,150 {\respwansites}.
We found that the  owners and websites are distributed across the globe, ranging respectively  over 29 and 47 different countries, including EU.
Out of these 670 websites, 52 (7.76\%) are in the EU.

\begin{figure}[]
    \centering
    \includegraphics[width=0.4\textwidth]{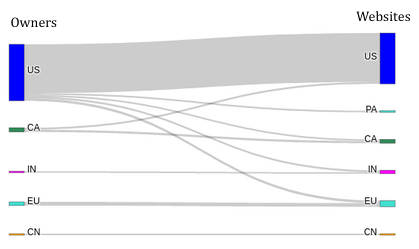}
    \caption{\textbf{{\Ourmethod} is geolocaly distributed.} \normalfont{{Corresponding countries of owners (left) and {\respwansites} (right) of respawned cookies.  We present the top 10 (owner,website)  geolocation. "EU" label represents the 27 member states of the EU.}}}
    \label{fig:geo}
\end{figure}

Figure~\ref{fig:geo} presents the registration countries of respawned cookies owners and websites where they are set.
We observe that top countries
of both respawned cookies and {\respwansites}  are not 
in the EU:
 356 (24.98\%) of the respawned cookies are both originated and included by domains from the US.
We also observed that respawned cookies on Chinese websites are only set by Chinese owners, and interestingly, websites registered in Panama are active in respawning as well (22 (3.28\%) of the studied 670 websites including respawning are from Panama).

\noindent
{{\bf Summary. }
{\Ourmethod} is commonly used: 5.95\% of \textit{General News} websites contain at least one respawned cookie.
We found that cookies are respawned in sensitive \textit{adult} websites as well, which leads to 
serious privacy implications: 
{\Ourmethod} is distributed across the globe, however, 
only 7.76\% of the websites that include respawning are in the EU.
Nevertheless, both EU and non-EU websites must comply with the GDPR 
as it is applicable independently of the country of registration of the website where 
EU users are monitored.}

\subsection{Tracking consequences of respawning}
\label{sec:tracking}
\IF{\#134 A: respawned cookies for cross-site tracking}
\subsubsection{Persistent cross-site tracking with respawned cookies}
Basic tracking via third-party cookies~\cite{Roes-etal-12-NSDI,FOUA-PETS-20} is the most known
tracking technique that allows third parties to track 
users across websites, hence to recreate her browsing history. 
When a third party cookie that enables cross-site tracking is respawned, such tracking becomes {\em persistent}.
That is, in contrast to regular third-party tracking, 
the user can not prevent 
it by deleting cookies.
Hence, respawned cookies enable persistent tracking 
that allows trackers to 
create larger users' profiles by linking users activity 
before and after 
they clean their browser.
Since the host is the domain to whom browser automatically sends the cookies,
we focus on the cookie \emph{host} and not on cookie owner.

Third party cookies allow trackers to 
track users \emph{cross-websites}~\cite{Roes-etal-12-NSDI}.
In this section, we 
only analyse third-party respawned cookies that can be used to track users across websites.
Note that all extracted respawned cookies are user specific (Section~\ref{sec:specific}) 
and therefore can be considered as {\em unique identifiers}. 
Out of {\respcookies} respawned cookies, 
528 (37.05\%) are third-party cookies.
In total, we identified 144 
unique hosts that have access to these cookies.
\begin{figure}[]
    \centering
    \includegraphics[width=0.4\textwidth]{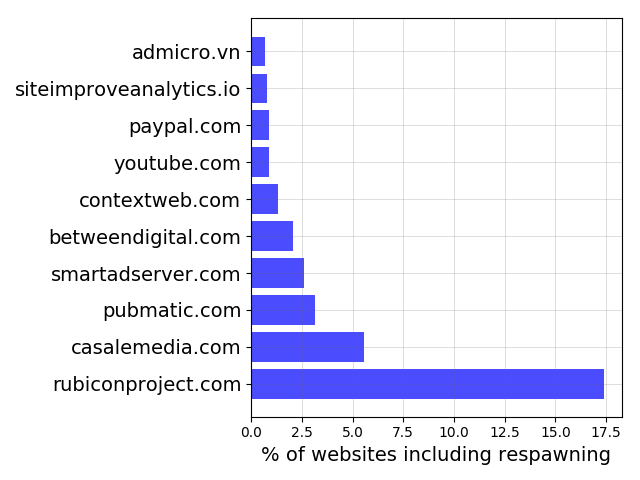}
    \caption{\textbf{Persistent third-party tracking based on respawned cookies.} \normalfont{Top 10 cross-site trackers using respawned cookies.}}
    \label{fig:basicT}
\end{figure}
Figure~\ref{fig:basicT} presents the top 10 
cross-site trackers that 
have access to respawned cookies.
We found that \texttt{rubiconproject.com} is the top domain: 
it has access to 
at least one respawned cookie on 200 (17.39\%) of the visited websites out of {\respwebsites}.
\texttt{Rubiconproject.com} defines itself as a publicly traded company, as it is automating the buying and selling of advertising~\cite{rubicon}.
 
\IF{\#134A: Lack of justification for cookie respawn for first parties}
 \subsubsection{{\Ourmethod} beyond deprecation of third-party cookies}
 \label{sec:fpTracking}
 Web browsers are moving towards deprecation of third party cookies which are the core of cross-site tracking~\cite{Sandbox}. 
\emph{Can this deprecation prevent cross-site tracking?}
In the following, we show how
{\ourmethod} can overcome browsers preventions.

 Via {\em persistent tracking with respawned cookies},
 domains can track users across websites without 
 third-party cookies. 
Consider the following scenario:  \texttt{example.com} and \texttt{news.com} include a fingerprinting script from  \texttt{tracker.com}. When the user visits these websites, the  script from \texttt{tracker.com} accesses the user's browser and machine features, and sets a corresponding first-party cookie.
 As a result, two first-party cookies are set in the user's browser 
and labeled with two different hosts:
\texttt{example.com} and \texttt{news.com}, but 
the values of these two cookies are identical, because they are created from the user's browser and machine features.
By respawning these two cookies on both websites, the owner \texttt{tracker.com} shows to be
able to track the user in a persistent way across sites with a
first-party cookie only.

We  analyzed the usage of the same (owner, key, value) first-party respawned cookie across different websites.
The {\respcookies} cookies correspond to 1,244 respawned (owner, key, value) instances,
out of which 
40 (3.21\%) are respawned on multiple websites  in a first party context with the same value (see Table~\ref{tab:crosstrackfp} in Appendix).
\texttt{wpbeaverbuilder.com}~\cite{Wpbeaverbuilder} is the top owner setting identical first party respawned cookies across websites.
It respawned the same cookie on 15 distinct websites. 
It defines itself as a WordPress page builder.
Its policy declare to collect user's information, but it does not precise the type of this information.
 
\noindent 
{{\bf Summary. }
{\Ourmethod} enable tracking 
across websites even when third party cookies are deprecated. 
We found 40 first party cookies that can serve for cross-site tracking.
}

\section{Is respawning legal?}
\label{sec:legal}

In this section, with a legal expert which is  a co-author, we evaluate the legal compliance of  {\respcookies} respawned cookies and reflect upon the 
applicability of current regulations in practice. 
Our legal analysis is based 
on the  General Data Protection Regulation (GDPR)~\cite{gdpr}
and the ePrivacy Directive (ePD)~\cite{ePD-09}, as well as in its recitals (which help legal interpretation of provisions in a specific context, but they are not mandatory for compliance).
The GDPR applies to the processing of personal data ~\cite{EDPB-4-07} and requires that companies need to choose a legal basis  to  lawfully  process  personal  data (Article 6(1)(a)).
The ePD provides \emph{supplementary} rules to the GDPR 
in particular in the  electronic communication sector, such as websites, and requires those, whether inside or outside the EU, to obtain \emph{consent} from users located in the EU for processing of their personal data.
We have additionally consulted the guidelines of both the European Data Protection Board (an EU advisory board on data protection, representing the regulators of each EU member state)~\cite{29WP} and the European Data Protection Supervisor (EDPS, the EU's independent data protection institution)~\cite{EDPS2016}. While these guidelines are not enforceable, they are part of the 
EU framework for data protection which we apply in this work to discern whether respawning is compliant.

To assess the legal consequences of respawning, the legal expert 
analysed 
legal sources 
to interpret cookie deletion.
To our surprise, we found  that there is 
\emph{no explicit legal interpretation of cookie deletion.}
Only the EDPS~\cite[Section 4.3.4]{EDPS2016}
noted that \textit{"if cookies requiring consent have disappeared, this is most probably because the user
deleted them and wanted to withdraw consent"}.
As a result, {\em cookie respawning also  does not have a clear legal interpretation} and merits attention for its plausible legal consequences. These consequences can arguably be derived, not only from the consent perspective, but also from the core principles of data protection, as discussed  in the following sections (fairness, transparency and lawfulness principles).
Thus, 
 owners of respawned cookies and website owners that embed those may be jointly responsible for their usage (Article 26~\cite{gdpr}) and may then be subject to fines of up to 20 million EUR (or 4\% of the total worldwide annual turnover of the preceding financial year,  Article 83(5)\cite{gdpr}).

\noindent
\subsection{Fairness Principle.} 
This principle requires personal
data to be processed fairly (Article 5(1)(a)). 
It requires that 
i) {\em legitimate expectations of users are respected} at the time and context of data collection, and
ii) there are no “surprising effects” or potential negative consequences occurring in the processing of user's data.


\noindent
\textbf{Findings: }
We consider that \emph{all {\respcookies} respawned cookies plausibly violate the fairness principle}, as respawning seems to be inconsistent with the user's expectations regarding respawned cookies after its deletion from her browser, and also considering the cookie's duration.

\noindent
\textbf{Suggestions for policymakers:}
It's hard to operationalize the high-level fairness principle into concrete requirements for  website owners and map it into legitimate expectations of users. 
Policy makers need to provide more concrete guidelines on the operationalization of this principle in the Web.



\noindent
\subsection{Transparency principle.}  
\label{sec:transp}
Personal data processing must be handled in a transparent manner in relation to the user (Article 5(1)(a)). This principle presents certain obligations for websites: 
i) inform about the \emph{scope and consequences} ~\cite{Transparency29WP} and the \emph{risks} in relation to the processing of personal data (Recital 39); 
ii) inform about the purposes, legal basis, etc. before processing starts (as listed in Art. 13);
iii) 
provide the above information in a concise, transparent, intelligible and easily accessible form 
(Art. 12).

\noindent
\textbf{Findings: }
We analyzed the privacy policies of the 10 top popular respawned cookie owners:
\textit{rubiconproject.com~\cite{rubicon},
casalemedia.com~\cite{casalmedia}, pubmatic.com~\cite{pubmaticpolicy}, adobedtm.com~\cite{adobedtm}
smartadserver.com~\cite{smartadserver}, bizible.com~\cite{bizible}, betweendigital.com~\cite{betweendigital}, maricopa.gov~\cite{maricopa}, wpbeaverbuilder.com~\cite{Wpbeaverbuilder}, and contextweb.com} (Figure~\ref{fig:owners}).
Some 
policies~\cite{rubicon,casalmedia,adobedtm,smartadserver,bizible} refer to the use of browser's features without referencing the consequences or risks thereof. 
Also, none of the policies refer to 
cookie respawning. As such, these seem to be in breach of the transparency principle.

\noindent
\textbf{Suggestions for policymakers:}
In practice, the description of data (purposes, legal basis, types of personal data collected, features used and its consequences) is often mixed within the text, which makes harder to extract concrete information therefrom~\cite{fou-IWPE-2020}.
Policy-makers 
need to converge on harmonized requirements and standard format for privacy policies.

\noindent
\subsection{Lawfulness Principle.} 
The ePD requires websites to obtain user consent to lawfully process personal data using
cookies.
When a cookie recreates itself without consent, every data processed henceforth could be considered unlawful due to lack of legal basis for personal data processing~\cite{EDPB-15-11}.
This practice incurs in violation with the lawfulness principle (Articles 5(1)(a) and 6(1) of the GDPR, and 5(3) of the ePD).
The EDPS~\cite{EDPS2016} already advised against the use of cookie respawning if the processing relies on users' consent.
It mentions that 
\textit{"cookie respawning would circumvent
the user’s will. In this case (...) institutions must collect again user’s consent". }

To evaluate compliance with the lawfulness principle, we need first to 
evaluate whether cookies are exempted or subject to consent.
The 29WP~\cite{EDPB-4-12} asserts that \textit{"it is the purpose 
that must be used to determine whether or not a cookie can be exempted from consent}".

Given that only a small percentage of cookies include  a description of their purposes~\cite{fou-IWPE-2020},
we adopted the Cookiepedia open database~\cite{cookiepedia} which has over 11 million cookies used across 300,000 websites and has been used in prior work~\cite{Urb-www-20}.
It uses the classification system developed by "The UK International Chamber of Commerce" (ICC) and relies on four common purposes of cookies: 
i) Strictly Necessary (which includes authentication and user-security);
ii) Performance (also known as analytics, statistics or measurement cookies);
iii) Functionality (includes customization, multimedia content); and iv) Targeting (known as advertising).
Even though this classification is not binding, we point to the fact that it is the largest database of pre-categorized cookies.

The 29WP~\cite{EDPB-4-12} adds two other characteristics contributing to determine whether cookies are exempted or subject to consent: duration (\emph{session and persistent cookies}) and context (\emph{first and third-party cookies}).
Building on the analysis made by Santos et al.~\cite[Table 5]{Sant-etal-20-TechReg} on the list of \emph{purposes} that are subject to consent and those that are exempted therefrom, we firstly studied the Cookiepedia purposes and then we derived which are the purposes subject to consent according to their duration and context. 
Table~\ref{tab:direct_consent} summarizes the Cookiepedia purposes requiring consent depending on the duration and  the context on which it is running.

\begin{table}[]
    \centering
    \begin{tabular}{|p{.8 cm}|m {3 cm}|m {3 cm}|}
        \hline
    &  Session  &  Persistent \\ \hline
    
   \makecell{First- \\ party}
        &
        

        Targeting/ Advertising 

         & 

       Targeting/ Advertising 
      
    \\ \hline

    \multirow{ 4}{*}{\makecell{Third- \\party }}     
 &  Targeting/ Advertising &     Targeting/ Advertising \\
& Performance & Performance  \\ 
&        Strictly necessary  &  Strictly necessary \\
& &  Functionality

      \\ \hline
    \end{tabular}
    \caption{\textbf{Purposes of Cookiepedia~\cite{cookiepedia} that require consent according to their context and duration.}}
    \label{tab:direct_consent}
         \vspace{-.5 cm}

\end{table}

\noindent
\textbf{Findings:}
In our study we crawled websites and even if a website provided a consent banner, we did not give consent thereto.
We evaluated whether respawned cookies are subject to or exempted from consent (as described in Table~\ref{tab:direct_consent}). 
As a result of our evaluation, we found that out of 336 respawned cookies categorized by Cookiepedia, 130 (38.69\%) are subject to consent.
Hence, these 130 cookies are in breach of the lawfulness principle.

\noindent
\textbf{Suggestions for policymakers:}
Companies can embed respawning and still claim respawned cookies are exempted of consent.
We analysed that both the duration and context of cookies contribute to determine whether cookies are exempted or subject to consent.
However, from a technical point of view, these criteria can be bypassed by domains that embed respawning.
As per \emph{duration}, session cookies can get recreated even after their elimination by the user.
Functionality cookies 
are exempted of consent when used as session cookies and are subject to consent when used in a persistent way~\cite{EDPB-4-12}. 
When respawned, such cookies can be used for a longer duration than previously envisaged.
We found that out of {\respcookies} respawned cookies, 446 (31.30\%) are session cookies.
Regarding \emph{context}, performance cookies  are exempted of consent when used in a first party context and are subject to consent when used as third party cookies. 
However, in practice,
a cookie set in the first party context can be considered as a third party cookie
in a context of a different website.
 We found that $4$ respawned cookies (host,key,value) appear as first- and third-party in different websites. These cookies are respectively set by \texttt{pornhub.com}, \texttt{mheducation.com}, \texttt{hujiang.com} and \texttt{fandom.com}.
Given that a cookie context and duration can be altered, 
these should not be used as a criteria  to evaluate the need of consent. 

\section{Conclusion}
\label{sec:conclusion}
This work presents
a large scale study of {\ourmethod}, a tracking technique that is
devoid of 
a clear legal interpretation in the EU legal framework. 
We employed a novel methodology to reveal the prevalence of {\ourmethod} in the wild.
The detection of such behavior and the identification of responsible domains can prove to be hard to achieve,
which impacts both the ability to block such behavior, and its legal assessment.
We believe this work can serve as a foundation for improvement of future regulation and protection mechanisms.

\bibliographystyle{plain} 
\bibliography{papers}
\clearpage
\clearpage
\section{Appendix}
\renewcommand\thesubsection{\Alph{subsection}}

\subsection{Machines characteristics}
\label{sec:appendChara}

Table \ref{tab:characteristics} presents the 
characteristics of machine A and machine B used in our study.

\begin{table}[!h]
    \centering
    \begin{tabular}{|p{2.4 cm}|p{2 cm}|p{2 cm}|}
    \hline
    \textbf{Characteristics} & \textbf{Machine A} & \textbf{Machine B} \\ \hline
       Date of the crawl &  March 2021 &  March 2021 \\ \hline
     OS &  Fedora 25  &  Fedora 31 \\ \hline
     Firefox version &  68.0 &  45.0.1 \\ \hline
     Location & France & France \\ \hline
     IP address &193.51.X.X & 138.96.Y.Y \\ \hline
     OpenWPM version & v0.9.0 &  v0.7.0 \\ \hline
     Language & English (en\_US) & German (de\_DE) \\ \hline
     Time zone & CET &  AKST \\ \hline
     Geolocation & France & Alaska  \\ \hline
     Do not track & Null & True \\ \hline
    \end{tabular}
    \caption{Crawls Characteristics. \normalfont{\textit{All crawls were performed from machine A except \textit{\CMremote} that was done from machine B.} }}
    \label{tab:characteristics}
\end{table}{}

\subsection{Additional results}
Table~\ref{tab:crosstrackfp} presents the top first-party cookies respawned across websites.
This practice is studied in Section~\ref{sec:fpTracking}.

 \begin{table}[!htbp]
    \centering
    \begin{tabular}{|p{3.2cm}p{3cm}|}
    \hline
    \textbf{Owner} & \textbf{Occurrence} \\ \hline \hline
\texttt{wpbeaverbuilder.com}  & 15  \\ 
\texttt{clarip.com}  & 13  \\ 
\texttt{maricopa.gov}  &  9  \\ 
\texttt{google-analytics.com }  &   7  \\ 
\hline
    \end{tabular}
    \caption{Top first-party cookies respawned across websites. \normalfont{{Every line in the table represents a cookie, hence the same owner can appear on multiple lines. Occurrence: presents the number of websites where the instance (owner,key,value) was respawned. T0:}} }
    \label{tab:crosstrackfp}
\end{table}

\end{document}